# A study of the limits of imaging capability due to water scattering effects in underwater ghost imaging

Yuliang Li,[1,2] Mingliang Chen,[1,*] Jinquan Qi,[1,2] Chenjin Deng,[1] Longkun Du,[1,3] Zunwang Bo,[1] Chang Han,[4] Zhihua Mao,[4] Yan He,[5] Xuehui Shao,[6] and Shensheng Han[1,2,7]

[1]*Key Laboratory of Quantum Optics, Shanghai Institute of Optics and Fine Mechanics, Chinese Academy of Sciences, Shanghai 201800, China*
[2]*Center of Materials Science and Optoelectronics Engineering, University of Chinese Academy of Sciences, Beijing 100049, China*
[3]*Institute for Quantum Science and Technology, College of Science, National University of Defense Technology, Changsha 410073, China*
[4]*State Key Laboratory of Satellite Ocean Environment Dynamics, Second Institute of Oceanography, State Oceanic Administration, Hangzhou 310012, China*
[5]*Key Laboratory of Space Laser Communication and Detection Technology, Shanghai Institute of Optics and Fine Mechanics, Chinese Academy of Sciences, Shanghai 201800, China*
[6]*National Laboratory of Aerospace Intelligent Control Technology, Beijing, 100089,China*
[7]*Hangzhou Institute for Advanced Study, University of Chinese Academy of Sciences, Hangzhou 310024, China*
*\*cml2008@siom.ac.cn*

**Abstract:** Underwater ghost imaging is an effective means of underwater detection. In this paper, a theoretical and experimental study of underwater ghost imaging is carried out by combining the description of underwater optical field transmission with the inherent optical parameters of the water body. This paper utilizes the Wells model and the approximate S-S scattering phase function to create a model for optical transmission underwater. The second-order Glauber function of the optical field is then employed to analyze the scattering field's degradation during the transmission process. This analysis is used to evaluate the impact of the water body on ghost imaging. The simulation and experimental results verify that the proposed underwater ghost imaging model can better describe the degradation effect of water bodies on ghost imaging. A series of experiments comparing underwater ghost imaging at different detection distances are also carried out in this paper. In the experiments, cooperative targets can be imaged up to 65.2m (9.3AL, at attenuation coefficient $c=0.1426m^{-1}$ and the scattering coefficient $b=0.052m^{-1}$) and non-cooperative targets up to 41.2m (6.4AL, at $c=0.1569m^{-1}$ and $b=0.081m^{-1}$) . By equating the experimental maximum imaged attenuation length for cooperative targets to Jerlov-I water ($b=0.002m^{-1}$ and $a=0.046m^{-1}$), the system will have a maximum imaging distance of 193m. Underwater ghost imaging is expected to achieve longer-range imaging by optimizing the system emission energy and detection sensitivity.

## 1. Introduction

Underwater exploration is vital for areas such as marine resource exploration and underwater rescue. Obtaining underwater images is one of the most important ways for humans to understand the ocean. Currently, there are two main types of underwater detection techniques: optical imaging [1, 2] and acoustic imaging [3, 4], with underwater optical imaging having the advantage of high resolution, unlike acoustic imaging. There are two types of underwater optical imaging: scanning and non-scanning. Scanning optical imaging includes conventional laser point scanning imaging [5, 6], streak tube laser imaging [9, 10], and vortex optical imaging [16-17]; non-scanning optical imaging includes underwater distance selective imaging [7, 8], and optical polarization imaging [11-15]. Underwater distance selective imaging and optical polarization imaging require high system emission energy and relatively short imaging distances. Scanning optical imaging has a low resolution and cannot meet the imaging needs of small and moving targets underwater; among them, streak tube laser imaging has a large amount of noise which is not conducive to target identification; vortex light imaging can filter out scattered light but requires a complex modulation system. Ghost imaging is considered to have the characteristics of long detection distance, high imaging resolution, and strong anti-interference capability [18-21], which makes it gain more and more attention in underwater detection.

Underwater ghost imaging has been studied in a number of different directions. Several papers have validated the unique advantages of ghost imaging in the imaging of scattering media from different perspectives, including Gong's early application of ghost imaging to the detection of targets in turbid liquids and the validation of ghost imaging as being resistant to interference [22]; Le et al. investigated the effects of underwater turbidity and field of view on the computational ghost imaging and validated that ghost imaging can still achieve good results despite the failure of classical optical imaging methods [23]. The effect of turbidity and field of view on computational ghost imaging was investigated by Le et al. To address the effect of water body backscatter on ghost imaging, Chen et al. analyzed the system parameters and water body parameters affecting the imaging performance in pulsed laser ghost imaging according to the underwater lidar equation, so as to control the effect of backscatter [25]. To analyze the role of underwater turbulence on ghost imaging, Luo and Zhang developed their respective physical models for underwater turbulence ghost imaging [26,27]. In the study of underwater illumination speckle pattern optimization, Yang verified that the Hadama pattern can effectively improve the quality of underwater ghost imaging [28]. In terms of algorithms, differential ghost imaging [24], binary method ghost imaging [29], and neural networks [30] are some of the schemes that can effectively improve the imaging quality of underwater ghost imaging.

The imaging capability limits of underwater ghost imaging are an important metric to follow in underwater imaging research. Previous studies have investigated the limits of the imaging capability of underwater ghost imaging systems mainly in terms of energy and signal-to-noise ratio. Water is a typical light absorbing and scattering medium, and for ghost imaging, in addition to the energy attenuation caused by the absorption and scattering effects of the water body during light transmission underwater, the "smoothing" effect of the scattering of the water body on the intensity fluctuations of the light field can also limit the imaging distance and image quality of the ghost imaging system. In this paper, we use the Wells model and the approximate S-S scattering phase function related to the intrinsic optical parameters of the water body to construct an underwater optical transmission model and then use the second-order Glauber function of the optical field [34] to describe the degradation of the scattering field during the underwater transmission process, to analyze the effect of the water body on ghost imaging. The good imaging performance of underwater ghost imaging is verified experimentally, and the consistency between simulation and experiment is verified in terms of degradation of underwater scattering and ghost imaging results.

## 2. Theoretical analysis of underwater ghost imaging

In this paper, the effect of the inherent optical parameters (e.g. scattering coefficient b, absorption coefficient a) of a homogeneous stationary water body on ghost imaging is investigated; the effect of turbulence is temporarily excluded from the scope of this paper. The absorption coefficient is mainly caused by the absorption of light at specific wavelengths by water molecules and colored dissolved organic matter (e.g. chlorophyll) reducing the transmission energy of light [47] and shortening the detection distance; scattering is mainly caused by tiny particles in the water, dominated by Mie scattering [46, 47], which deflects light from its original propagation direction and causes degradation of the speckle field [33].

Ghost imaging requires only the collection of fluctuations in the relative intensity of the reflected light from a target in the form of a barrel probe. With sufficient detection energy, a static scattering medium in the return optical path does not affect the detection of relative intensity fluctuations by ghost imaging [22]. The effect of speckle field transmission from the emitting side to the object surface in the water medium on ghost imaging is the main study in this paper.

The principle diagram of underwater ghost imaging is shown in Figure 1. The pseudo-thermal light is generated by a laser beam passing through a rotating hair glass, which is split into a detection and a reference light path by a beam splitter. In the detection light path, the pattern is directed through the body of water onto the object, and the light reflected from the object is received by the bucket detector $I_{Bh}^{(i)}$ through the water medium. In the reference light path, the pattern propagates directly through the air to the CCD camera $I_r^{(i)}(x,y)$ (where $i = 1 \cdots M$ is the number of samples).

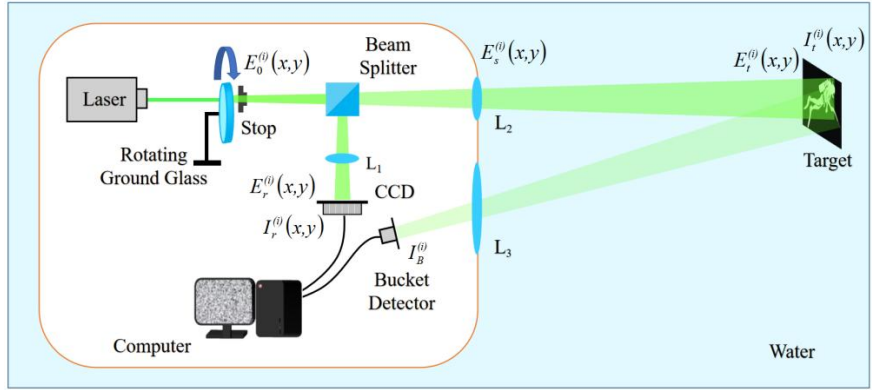

Fig. 1. Principle diagram of underwater pseudo-thermal light ghost imaging

The optical field at the target surface $E_t^{(i)}(x,y)$ is related to the optical field at the emission surface of the ground glass $E_0^{(i)}(x,y)$, the coherent transfer function of the detection optical path within the system $h_{t1}(x,y)$, and the coherent transport process in the water body between the target and the system. In this paper, the coherent transport process of the speckle pattern in the water body is expressed as the combination of coherent transport in the same refractive index medium without scattering effect and incoherent scattering diffusion in the water body. The speckle pattern intensity distribution of the target surface in water is expressed as:

$$\begin{aligned}
I_t^{(i)}(x,y) &= I_{t0}^{(i)}(x,y) \otimes h_w(x,y) = \left| E_{t0}^{(i)}(x,y) \right|^2 \otimes H_w(x,y) \\
&= \left| E_0^{(i)}(x,y) \otimes h_{t1}(x,y) \otimes h_{t2}(x,y) \right|^2 \otimes H_w(x,y) \\
&= \left| E_s^{(i)}(x,y) \otimes h_{t2}(x,y) \right|^2 \otimes H_w(x,y) \\
&= \iint_{-\infty}^{\infty} \left\{ \iint_{-\infty}^{\infty} dx_{t0} dy_{t0} \left| E_s^{(i)}(x,y) \exp\left[ \frac{j\pi}{\lambda z}\left((x - Mx_{t0})^2 + (y - My_{t0})^2\right) \right] \right|^2 \right\} dx_w dy_w \\
&\quad \times H_w(x - x_w, y - y_w)
\end{aligned} \qquad (1)$$

Where $\otimes$ is the convolution symbol, $E_s^{(i)}(x,y) = E_0^{(i)}(x,y) \otimes h_{t1}(x,y)$ is the system emitted optical field, $h_{t2}(x,y)$ is the coherent transfer function of a medium with the same refractive index as the water body without scattering effects, $H_w(x,y)$ is a incoherent point spread function for light transmission underwater, $z$ is the underwater transmission distance, $\lambda$ is the wavelength, and $M$ is the scattering magnification related to the refractive index of the water body and the system emission angle. As the static homogeneous scattering medium in the return optical path does not affect the detection of relative fluctuations in intensity by ghost imaging [22], the bucket detection value of the receiving system is proportional to the bucket detection value of the light field intensity at the target reflecting surface. The bucket detection value of the receiving system can be expressed as:

$$I_B^{(i)} = k \iint I_t^{(i)}(x,y) T(x,y) dxdy \tag{2}$$

where $k$ is a positive proportionality constant and $T(x,y)$ is the reflectance function of the target.

The optical field of the CCD imaging surface in the reference optical path $E_r^{(i)}(x,y)$ is the convolution of the transmitted optical field of the ground glass surface $E_0^{(i)}(x,y)$ with the coherent transfer function of the reference optical path within the system $h_r(x,y)$. The pattern intensity distribution recorded by the reference optical path CCD is expressed as:

$$I_r^{(i)}(x,y) = \left|E_r^{(i)}(x,y)\right|^2 = \left|E_0^{(i)}(x,y) \otimes h_r(x,y)\right|^2 \tag{3}$$

The internal optical path of the system is designed so that the target plane is conjugate to the reference optical path CCD imaging plane. Second-order correlation is calculated to obtain the target image:

$$O_{GI}(x,y) = \langle I_r^{(i)}(x,y) I_B^{(i)} \rangle - \langle I_r^{(i)}(x,y) \rangle \langle I_B^{(i)} \rangle \tag{4}$$

Where $\langle \bullet \rangle$ denotes the average, $O_{GI}(x,y)$ is the reconstructed underwater target image.

In order to evaluate the effect of pattern degradation caused by aqueous media on ghost imaging, the second-order mutual correlation (i.e., the normalized second-order Glauber function) is used in this paper to describe the irregularity of thermal light intensity fluctuations on the object surface [34]. The normalized second-order Glauber function for two points $(x_1, y_1, t_1)$, $(x_2, y_2, t_2)$ on the spacetime is defined as:

$$g^{(2)}(x,y) = \frac{\langle I_1(x_1,y_1,t_1) I_2(x_2,y_2,t_2) \rangle}{\langle I_1(x_1,y_1,t_1) \rangle \langle I_2(x_2,y_2,t_2) \rangle} \tag{5}$$

In this research, the level of speckle pattern degradation is obtained by using the above coherent propagation and point spread functions to simulate the autocorrelation of the speckle intensity distribution irradiated at the target plane at the same moment ($t_1 = t_2$). The maximum value of $g^{(2)}(x,y)$ is used to represent the visibility of ghost imaging and the full width at half maxima (FWHM) of $g^{(2)}(x,y)$ is used to characterize the degree of degradation of underwater ghost imaging resolution. For pseudo-thermal light fields, the maximum theoretical value of $g^{(2)}(x,y)$ is 2. The higher the value, the better the visibility of the ghost image. When it degrades to 1, the correlation properties are completely destroyed and imaging is not possible. The effect of speckle pattern degradation on imaging is assessed by the normalized Glauber function of the speckle center point and other points:

$$g^{(2)} = \frac{\langle I_t^{(i)}(x,y) I_t^{(i)}(x_0,y_0) \rangle}{\langle I_t^{(i)}(x,y) \rangle \langle I_t^{(i)}(x_0,y_0) \rangle} \quad (6)$$

where $I_t^{(i)}(x,y)$ is the intensity value of each spatial point of the speckle pattern irradiated on the surface of the target and $I_t^{(i)}(x_0,y_0)$ is the intensity value of the central pixel of the speckle pattern irradiated on the target surface.

The scattering and absorption effects of water make the beam spread and intensity diminish. In order to quantitatively describe the transmission of light in stable homogeneous water, many scholars describe it as a point spread function (PSF) or modulation transfer function (MTF) related to the inherent optical properties (IOPs) (scattering phase function $s(\theta)$, absorption coefficient $c$, scattering coefficient $b$ in the water) and transmission distance $z$ [36]. Duntley, Wells, and Dolin have obtained their respective PSF or MTF models by experimental or theoretical derivation[37, 31, 38]. Wells derived a modulation transfer function (MTF) model in the frequency domain from the firstness principle [31], which can accurately describe the transmission process of an underwater scattered beam in a small-angle approximation, and the correctness of his model was verified in [39] and other papers. In this paper, the Sahu-Shanmugam (S-S) scattering phase function proposed in the literature is linearly approximated in a small angular range (0.1° to 5°) in a logarithmic coordinate system [33]. Then the scattering phase function is Hankel transformed and substituted into Wells' MTF model to obtain the modulation transfer function of the pattern propagation in water. MTF is the frequency domain expression of PSF, which can describe the degradation effect of seawater on the beam.

The approximate derivation of the scattering phase function is obtained from Appendix 1. The normalized scattering phase function is Hankel transformed to obtain:

$$\widetilde{S}(\psi) = 10^q [\Gamma(-p/2)]^{-2} B^{-1} \pi^{-p} \cdot [\sin(-p\pi/2)]^{-1} (\psi)^{-p-2} \quad (7)$$

where $\Gamma(u) = \int_0^\infty e^{-t} t^{u-1} dt$ is the second type of Euler integral function.

Substituting equation (7) into the Wells model:

$$MTF = F(\psi) = \exp\left[-cz + b \int_0^z \widetilde{S}\left(\frac{\psi r}{z}\right) dr\right] \quad (8)$$

The modulation transfer function MTF obtained by simplification is expressed as:

$$F(\psi) = \exp\left\{-cz - bz \left[\frac{10^q (\Gamma(-p/2))^{-2}}{B \cdot \pi^p \cdot (p+1) \cdot \sin(-p\pi/2)}\right] \cdot \psi^{-p-2}\right\} \quad (9)$$

where $\psi$ is spatial angular frequency. The MTF formula in the frequency domain shows that it is essentially a low-pass filter, and its effective band-width decreases as the target distance increases, thus increasing the spatial blurring of the image. When b→0, MTF→exp(-az) in the frequency domain is a constant independent of the spatial angular frequency; at the same time, when z→0, MTF→1. A constant in the frequency domain, is transformed to the spatial domain as an impulse response function, and the physical reality indicates that the pattern is not degraded. The above modulation transfer function is Inverse Hankel transformed to obtain the point spread function:

$$H_w(\theta) = HT^{-1}[F(\psi)] \quad (10)$$

where $HT^{-1}$ denotes the inverse Hankel transform. The scattering spread of the beam is considered to be centrosymmetric, then the scattering angle is denoted as $\theta = \sqrt{x^2+y^2}/z$ and $(x,y)$ are the coordinates of the spatial position of the pattern irradiation in the target plane.

Therefore, the point spread function can be expressed as $H_w(x, y)$ in the spatial coordinate. Substituting equation (10) into equation (1) and equation (6), the simulation gives the maximum and FWHM of $g^{(2)}(x, y)$ variation curves for different distances and different water body parameters.

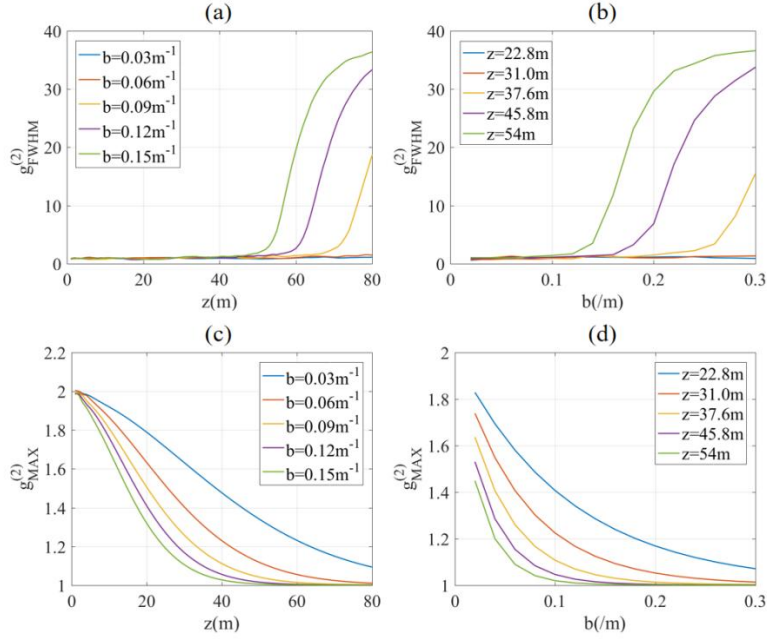

Fig. 2. Comparing the trend of $g^{(2)}_{MAX}$ and $g^{(2)}_{FWHM}$ at different scattering coefficients b or different distances z. (a) Plots of the FWHM of $g^{(2)}(x, y)$ with distance z; (b) Plots of the FWHM of $g^{(2)}(x, y)$ with scattering coefficient b; (c) Plots of the maximum value of $g^{(2)}(x, y)$ with distance z; (d) Plots of the maximum value of $g^{(2)}(x, y)$ with scattering coefficient b.

The scattering phase function parameters in the simulation are taken as the condition of $B_p = 0.0183$. As shown in Fig. 2(b) and Fig. 2(d), for different distances (z = 22.8m, 31m, 37.6m, 45.8m, 54m) and scattering albedo $\omega = b/c = 0.4$, $g^{(2)}_{MAX}$ and $g^{(2)}_{FWHM}$ decrease with increasing scattering coefficient b; Figures 2(a) and 2(c) show the decrease of $g^{(2)}_{MAX}$ and $g^{(2)}_{FWHM}$ with increasing distance z for albedos $\omega = 0.4$ and different scattering coefficients (b = 0.03/m, 0.06/m, 0.09/m, 0.12/m, 0.15/m). The simulation results can reflect the effect of water scattering on pseudo-thermal optical ghost imaging.

## 3. Experimental research

The experimental device consists of a transmitting modulation system, receiving system, control system, sealed housing, etc., using laser irradiation rotating ground glass to produce the intensity fluctuation of the patterns. Combining the characteristics of pulsed laser detection, the bucket detector gating technique is used to get the reflected light intensity of the target. The receiving system is placed separately from the transmitting system, and the transmitting system and receiving system are located in two layers of the sealed box. The above two settings reduce the impact of backward scattering as much as possible. To reduce the absorption effect of the water, the device uses a pulsed laser of 532 nm wavelength to irradiate the rotating rough glass to produce pseudo-thermal light. The single pulsed light energy of the system output is 37uJ and the receiving bucket detector is PMT. The PMT uses the photoelectric sensor module H11526 series of HAMAMATSU company with a gating function. the diagram of the transmitting and receiving system is shown in Figure 3. The CCD in the transmitting system collects the reference arm pattern intensity distribution, and the PMT in the receiving system collects the object arm bucket detection values reflected from the target. Monitoring CCD in the receiving optical system for adjusting the position of the receiving target in the field of view. The underwater emission angle of the system is $0.43°$. The pattern size at the emission end is $0.003 \times 0.003$ m. Both the reference arm pattern and the resulting image are $256 \times 256$ pixels.

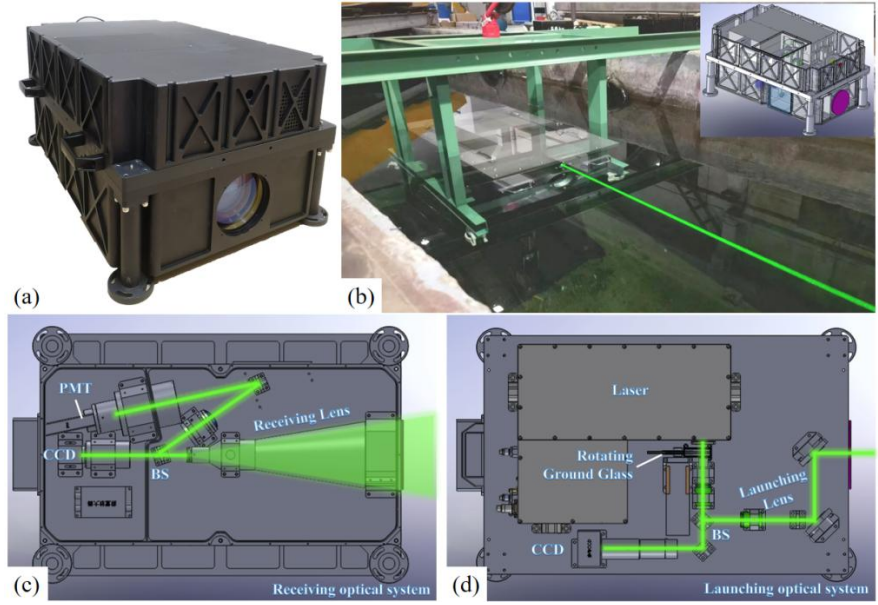

Fig. 3. **System transmitting and receiving system device diagram.** (a) Optical system package drawing; (b) Underwater experimental seal; (c) Receiving optical system; (d) Transmitting optical system. The receiving system is placed separately from the transmitting system, and the transmitting system and receiving system are located in two layers of the sealed box. The receiving system uses a pulsed laser of 532 nm wavelength to irradiate the rotating rough glass to produce pseudo-thermal light. The reference arm CCD records patterns of the reference arm. The PMT in the receiving system collects the object arm bucket detection values reflected from the target. The monitor CCD in the receiving system is used to adjust the position of the objects in the field of view.

The underwater environment for the experiments is a large ship-towing pond, which is rich in algae and suspended microscopic particles that play a very important role in the absorption and scattering of light. The scattering and absorption coefficients of the water body were measured using an AC-S hyperspectral meter from WET Labs, USA. This instrument measures the attenuation coefficients $c$ and absorption coefficients $a$ for each wavelength of light in the water and calculates the scattering coefficient $b = c - a$. According to the Beer-Lambert Law, the attenuation effect of light in water can be defined as $I = I_0 e^{-cz}$, $I$ and $I_0$ as the light intensity after the propagation distance z and the light intensity at the light source respectively, and $cz = -\ln(I/I_0) = 1$ is defined as 1 Attenuation

Length (AL), which represents the propagation distance at which the light intensity is attenuated $1/e = 0.3679$ times.

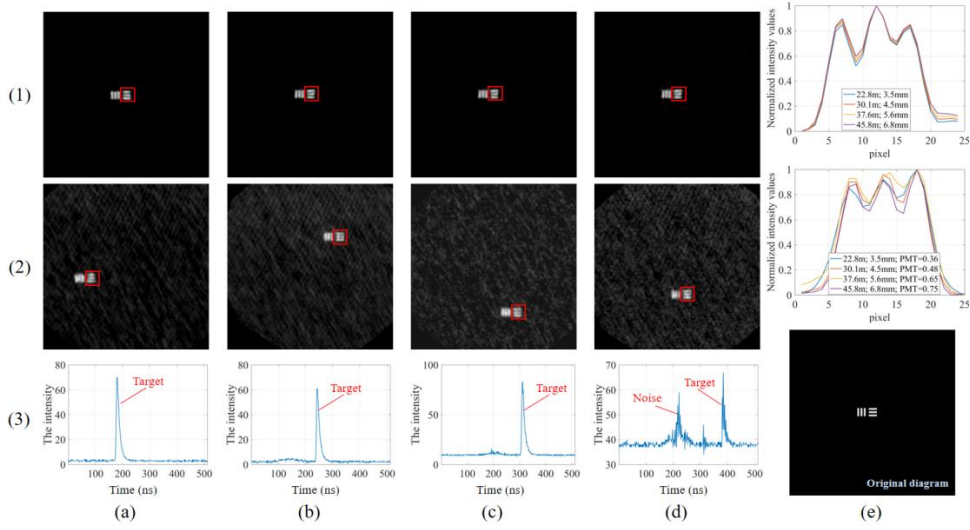

Fig. 4. **Comparison of experimental and simulation results.** (1) is obtained from the simulation, (2) is obtained from the experiment, and (3) is the echo curve obtained from the PMT acquired in the experiment. Figure (3) (e) is the original diagram "three slits" used for the simulation. The parameters of simulation are the same as the experimental. (a), (b), (c), and (d) are the imaging results for targets of different sizes at different distances (making the imaging scales the same). The target slit spacing of the "three slits" used in (a), (b), (c) and (d) are 3.5mm, 4.5mm, 5.6mm and 6.8mm; the underwater distances from the target to the system are 22.8m, 30.1m, 37.6m and 45.8m; the PMT detector gains are 0.36, 0.48, 0.65 and 0.75 respectively. The scattering coefficients b of the water bodies measured in experiments (a) ~ (d) were 0.081, 0.066, 0.066, and 0.061; and the absorption coefficients a were 0.076, 0.079, 0.079, and 0.061, respectively.

In order to more specifically characterize the effect of the aqueous medium on imaging quality, underwater pseudo-thermal optical ghost imaging simulations and the experiments with high-reflective targets underwater of the corresponding size were carried out for the three-slit image shown in Figure 4. In order to verify the theory presented in this paper, experiments and simulations were carried out under the same conditions: the emission angle of the system is $0.43°$; the pattern size at the emission end is $3\text{cm} \times 3\text{cm}$; the reference arm pattern and the resulting image are both $256 \times 256$ pixels.

Because the emission angle is fixed, we ensure that the imaging results of the objects have the same scale by changing the target distance and size. As shown in Figure 4, both the simulation and the experiment use the original GI algorithm. The difference is that the simulation does not incorporate the backscattering noise, while the experiment cannot be removed by the original GI algorithm due to the backscattering noise near the target. The results of simulation and experimental reconstruction were compared as shown in Figure 2 and Figure 4, it can be seen that the decreasing trend of resolution and the speckle degradation of the detection arm under the water is similar, and the experimental and MTF theoretical simulations are consistent.

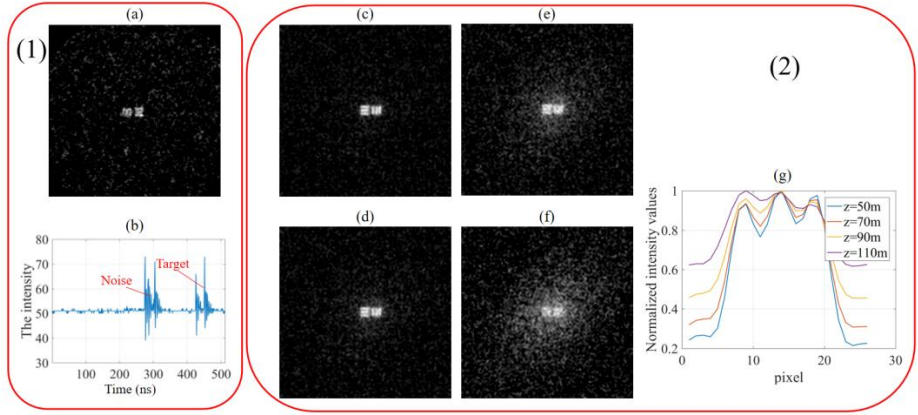

Fig. 5. (1) is the experimental results. (a) is second-order correlation calculation results for a three-slit target with a slit width of 8 mm at a distance of 54 m underwater with a PMT gain of 0.9. (b) is the PMT echo curve of experiment. (2) is obtained from the simulation. (c), (d), (e) and (f) are the simulation results at 50m, 70m, 90m and 110m underwater distances. (g) is a graph of the resolution curve under the condition of simulating different distances z.

Figures 5(a) and 5(b) show the experimental results, as shown in the figure the echo energy of the target at 54m is weak and the signal-to-noise ratio is poor. To show more specifically the effect of the water on ghost imaging, simulations were used for imaging at longer distances Figure 5 (c-g), with $b = 0.06\text{m}^{-1}$ and $\omega = b/c = 0.4$, for the simulation. From the simulation results, the resolution of the target gradually decreases with increasing distance at the same angular resolution. The target is resolved exactly at three slits at z=70m from the system (refer to Rayleigh criterion: the intensity value at the depression is 81% of the intensity value at the peak). The simulation results show that a system with an underwater angular resolution of 0.15mrad can remain resolvable for 10.5 attenuation lengths if the energy emitted is sufficiently large.

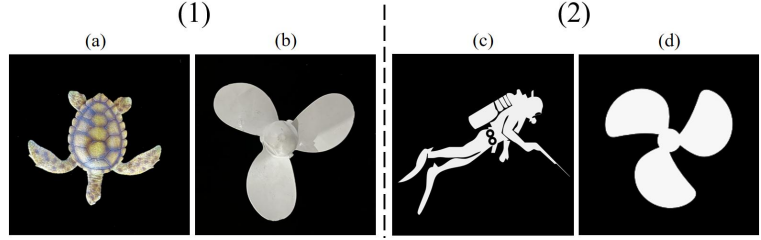

Fig. 6. **Actual targets and high reflectivity targets.** (1) is the non-collaborative targets, where (a) is "Turtle" and (b) is "White Propeller". (2) is the collaborative targets, where (c) is "Frogman" and (d) is "Highly Reflective Propeller".

In order to investigate the effect of different targets and different algorithms on the maximum detection distance of this system, image reconstruction was applied with the GI, DGI[24], and the TVAL3[35] algorithm, and two non-collaborative targets ("Turtle" and "White Propeller") and two collaborative targets ("Frogman" and "Highly Reflective Propeller") are used to verify the longest distance that can be imaged by this system. The target is shown in Figure 6. The reflectivity of the highly reflective targets is 0.5. The size of the "Turtle" is $12\text{cm} \times 13\text{cm}$; the diameter of the "White Propeller" is 20cm; the length of the "Frogman" is 43cm; and the diameter of the "Highly Reflective Propeller" is 31cm.

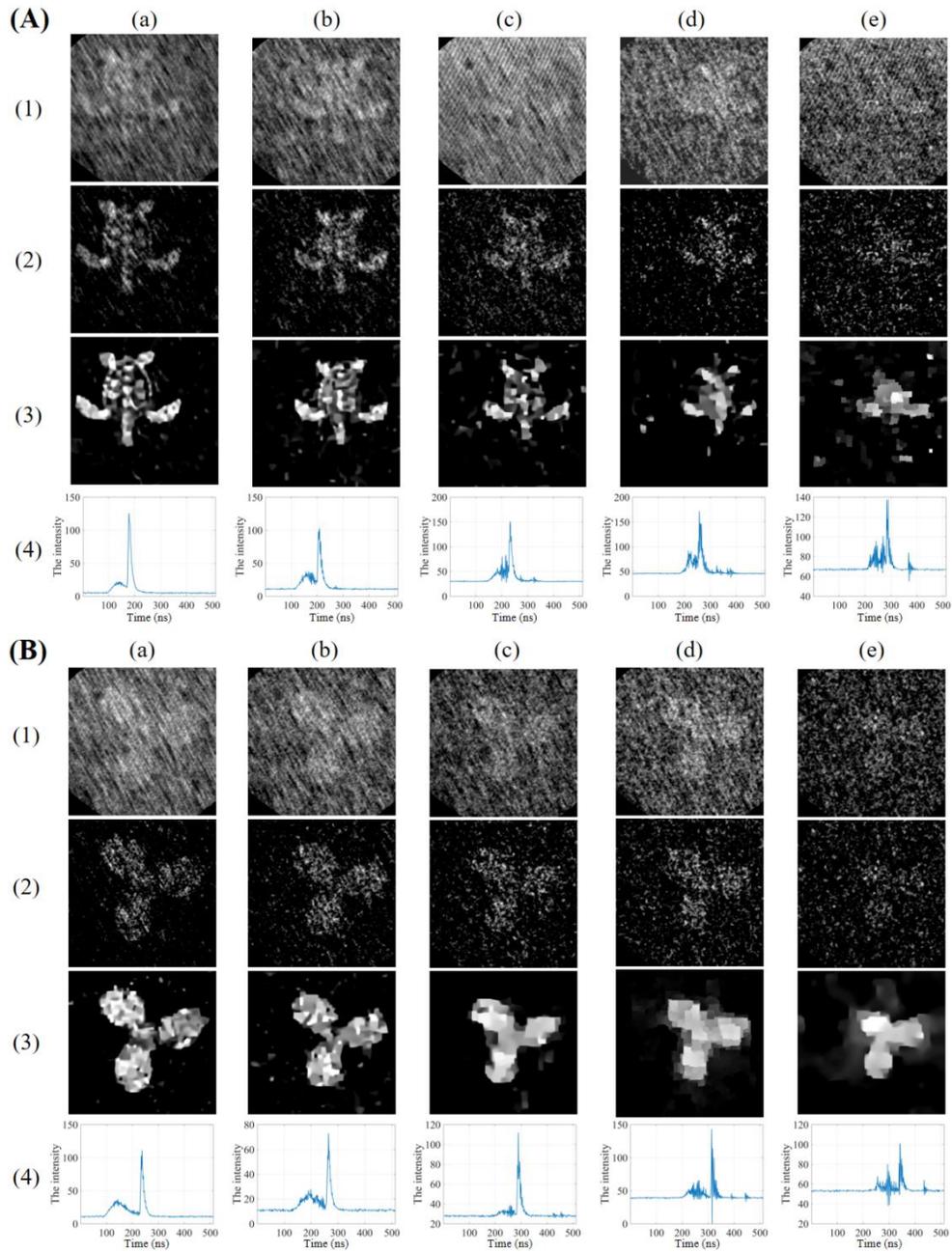

Fig. 7. **"Turtle" and "White Propeller" imaging results at different distances.** (1), (2) and (3) are the reconstruction results of GI algorithm, DGI algorithm, and TVAL3 algorithm, respectively. (4) is the curve of the echo signals. Figure (A) shows the imaging results of the "Turtle" target, the imaging distances from (a) to (e) are 22.8m, 25.8m, 28.8m, 31.8m and 34.8m respectively; the corresponding PMT gains are 0.55, 0.62, 0.75, 0.82 and 0.9. Figure (B) shows the imaging results of the "White Propeller" target, the imaging distances from (a) to (e) are 29.1m,

32.2m, 35.2m, 38.0m and 41.2m respectively; the corresponding PMT gains are 0.60, 0.65, 0.72, 0.75 and 0.9.

As shown in the Figure 7, as the distance increases, the PMT detector gating time is adjusted to move toward the target position so that the backscatter is as much as possible outside the gating range. However, as the PMT gain increases, the sensor noise also increases, and the echo signal-to-noise ratio at the target position deteriorates. As shown in Figure 7, the actual distance of the target from the system and the PMT gain have been marked in the echo curve. In the imaging experiment of the non-cooperative target, the scattering coefficient $b=0.081m^{-1}$ and the attenuation coefficient $c=0.1569m^{-1}$ were measured. The images show that the "Turtle" was imaged at a distance of about 34.9 m. The "White propeller" had a greater reflectivity and was imaged at a distance of about 41.2 m. The imaging distances for the non-cooperative targets "Turtle" and "White Propeller" were calculated to be 5.46 AL and 6.4 AL respectively.

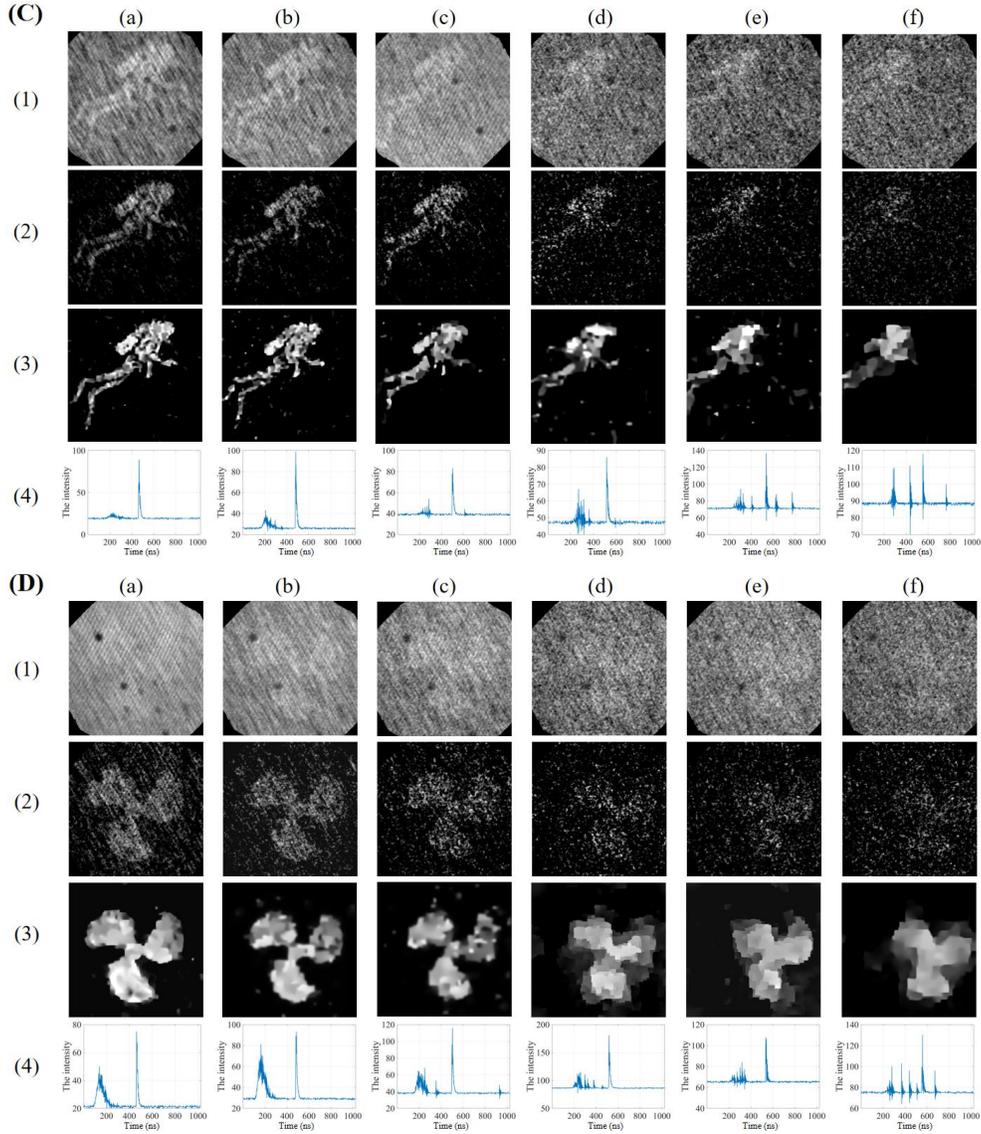

Fig. 8. **"Frogman" and "Highly Reflective Propeller" imaging results at different distances.** (1), (2) and (3) are the reconstruction results of GI algorithm, DGI algorithm, and TVAL3 algorithm, respectively. (4) is the curve of the echo signals. Figure (C) shows the imaging results of the "Frogman" target, the imaging distances from (a) to (e) are 55.1m, 57.0m, 59.1m, 61.1m, 63.1m, and 65.2m respectively; the corresponding PMT gains are 0.67, 0.72, 0.78, 0.83, 0.88 and 0.95. Figure (D) shows the imaging results of the "Highly Reflective Propeller" target, the imaging distances from (a) to (e) are 55.0m, 57.0m, 59.0m, 61.1m, 63.1m and 65.1m respectively; the corresponding PMT gains are 0.67, 0.72, 0.78, 0.83, 0.86 and 0.92.

As shown in Figure 8, the actual distance of the target from the system and the PMT gain have been marked in the echo curve. In the imaging experiment of the cooperative target, the scattering coefficient $b=0.052m^{-1}$ and the attenuation coefficient $c=0.1426m^{-1}$ were measured. The reflected energy of the two collaborative targets is similar, and the image quality decreases as the received signal-to-noise ratio decreases. The experimental cooperative target was imaged underwater at a distance of 65.2 m. Combining the optical parameters of the water column, the maximum attenuation length was calculated to be 9.3 AL. Equivalent to Jerlov-I water body conditions ($b=0.002m^{-1}$, $a=0.046m^{-1}$), the system would act at a maximum distance of 193 m. As can be seen from the PMT amplification gain variation and echo curves, the signal detected at the system's detection distance limit position contains a large amount of noise, resulting in eventual failure to image.

## 4. Discussion

1. Pseudo-thermal light ghost imaging images have a small field of view and high resolution. This requires the measurement and fitting of very small angular scattering phase functions (< $0.1°$). In this paper the field of view is $4.3°$, the image size is 256x256 pixels and the angle between pixels is $0.016°$. The instrumental accuracy of the measured scattering function is only $0.1°$. In this study, the volume scattering function (or scattering phase function) for less than $0.1°$ is fitted using a log-linear approximation in Cartesian coordinates, and there is room for improvement in its accuracy. More accurate PSF requires more accurate VSF, and the limitations of VSF measurement instruments do not allow for more accurate measurements of the volume scattering function. The study of instruments that measure the volume scattering function at very small angles will allow for more accurate PSF and facilitate the development of underwater high-resolution optical imaging.

2. In pulsed laser underwater ghost imaging, distance-selected slice imaging at the target is performed in such a way that the backscatter generated by the body of water within the distance-selected time range is picked up by the barrel detector. Backscatter noise is not discussed or studied in this paper. However, the effect of backscatter on image quality can be significant. The usual methods for reducing backscatter can be to reduce the selective pass time range or to filter out backscatter using other dimensions of light. Reducing the selective time width is very demanding on hardware such as lasers and can add significantly to the cost. Backscatter can be better filtered according to other dimensions of light, e.g. laser ghost imaging in coherent detection mode can use the optical coherence properties to filter out scattered light from the echo signal, promising higher signal-to-noise ratios and longer-range imaging results during underwater imaging.

3. Due to the different properties of the water and air media, the steep attenuation of energy is a problem that cannot be ignored in underwater optical detection. For practical systems, the maximum effective imaging (required resolution) distance is not only related to the degradation of the spot shining on the target as analyzed in this paper but also to the reflected light energy and detection signal-to-noise ratio of the target obtained by the detector. The energy and detection S/N ratio can be derived from the underwater lidar equations and are related to the actual transmitting power of the system, the transmitting angle, the receiving aperture, the angle between the transmitting and receiving axes, the reflective properties of the target and the inherent optical parameters of the water body (scattering coefficient, absorption coefficient, attenuation coefficient, body scattering function, backward scattering rate, etc.).

## 5. Conclusions

In this paper, the Wells model and an approximate S-S scattering phase function are used to describe the degrading effect of water on the spot transmission of the detection optical path. The normalized second-order Glauber function is also used to describe the effect of water on ghost imaging. In the experiments, the non-cooperative target imaging distance reaches 6.4 AL and the cooperative target imaging distance reaches 9.3 AL under the condition that the single pulse emission energy of the system is 37uJ. Equating the experimental longest achievable decay length of the cooperative target to the Jerlov-I water body condition, the longest action distance of the system will reach 193m. This paper provides a valid reference for long-range underwater pseudo-thermal optical ghost imaging.

## Appendix 1  Deduction and simulation of scattering phase function

Due to the small field of view of ghost imaging (the imaging field of view of the experimental device in this paper is 0.43°), this paper is only concerned with the scattering phase function matching accuracy in the small angle range. The S-S scattering phase function [33] is based on the Mie-scattering theory to approximate the polydisperse particle system (typical marine environment) to a monodisperse particle system to obtain the scattering phase function $s(\theta)$ fitted in the angle range of 0.1° to 5°:

$$\lg[s(\theta)] = P_1 \cdot [\ln(\theta)]^2 + P_2 \cdot \ln(\theta) + P_3 \quad (11)$$

where $\theta$ is the scattering angle and $P_i$ ($i$=1,2,3) is a function of the slope of the particle size distribution ($\xi$) and the real index of refraction of particles ($n$). $\xi$, $n$ and $P_i$ can be found when the backward scattering rate $B_P$ and the optical wavelength $\lambda$ are given, and the specific parameters are calculated by referring to the literature [33]. Referring to the method of Petzold [40] for the logarithmic linear extension of the phase function for very small angular scattering, the logarithmic function linear approximation of the S-S scattering phase function equation (11) in the range of $0.1°$ to $5°$ is presented in this paper as follows:

$$\lg[s(\theta)] = p \cdot \lg(\theta) + q \quad (12)$$

where $p$ is the slope of the linearly fitted log-log curve and $q$ is its intercept. And the range of $\theta$ definition of Eq. (12) is extended from 0° to 5°. Equation (11) and equation (12) are equal on both sides of the equal sign, that is, $P_1 \cdot [\ln(\theta)]^2 + P_2 \cdot \ln(\theta) + P_3 = p \cdot \lg(\theta) + q$. The approximation of Eq. (11) and Eq. (12) (taking $\ln(10)\ln(\theta) \approx -5$, $\ln(10) \approx 2.3$) is obtained by solving for the minimum root-mean-square error of the two function in the range from $0.1°$ to $5°$. The scattering phase function is obtained as:

$$s(\theta) = 10^{p \cdot \lg(\theta) + q} = \theta^p \cdot 10^q = \theta^{\ln(10) \cdot [P_1 \cdot \ln(\theta) + P_2]} \cdot 10^{P_3} \approx \theta^{-5P_1 + 2.3P_2} \cdot 10^{P_3} \quad (13)$$

Equation (13) is an approximation to Equation (11). The scattering phase function is a normalized bulk scattering function whose integral over the spherical solid angle is 1. We normalize the obtained scattering phase function by dividing its integral over the spherical solid angle from 0 to π:

$$\tilde{s}(\theta) = \frac{s(\theta)}{B} = 10^{P_3} B^{-1} \theta^{-5P_1 + 2.3P_2} \quad (14)$$

$$B = 2\pi \int_0^\pi s(\theta)\sin(\theta)d\theta = 2\pi \int_0^\pi \theta^{-4P_1 + 2.3P_2} \cdot 10^{P_3} \sin(\theta)d\theta \quad (15)$$

When the optical properties parameters of the water body are constant, $B$ is the normalization constant. The normalized scattering phase function is Hankel transformed to obtain:

$$\widetilde{S}(\psi) = 10^q [\Gamma(-p/2)]^{-2} B^{-1} \pi^{-p} \cdot [\sin(-p\pi/2)]^{-1} (\psi)^{-p-2} \qquad (16)$$

where $\Gamma(u) = \int_0^\infty e^{-t} t^{u-1} dt$ is the second type of Euler integral function.

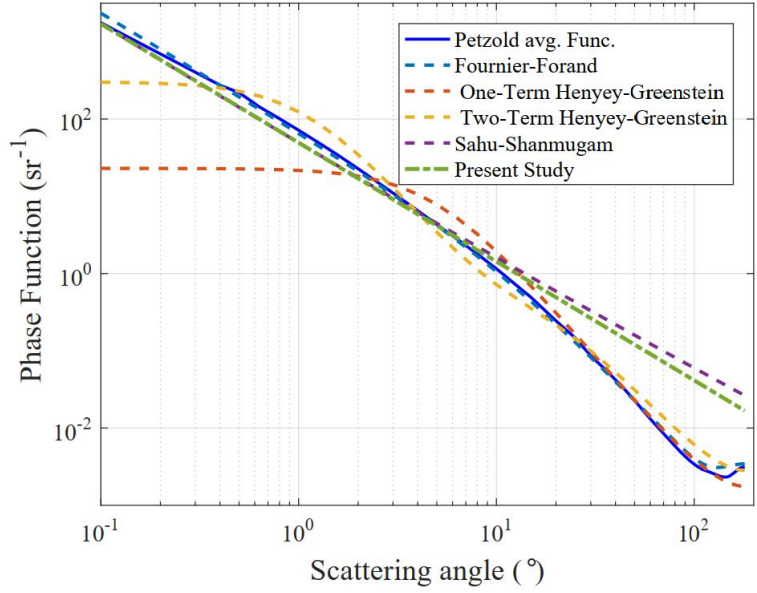

Fig. 9. Comparison of the phase functions obtained through different models for the scattering angles from 0.1° to 180° to emphasize the large deviation of different phase functions.

All phase functions are calculated for the corresponding backscattering ratio of $B_p = 0.183$ which is the $B_p$ value for Petzold's average particle phase function taken as the reference phase function for the present study. The curves for each phase function from 0.1° to 180° are shown in Fig. 9, but this paper is only concerned with the approximation of the small angle range. By defining the root mean square error (RMSE) rate to compare the approximate linear logarithmic phase function, the Sahu-Shanmugam (S-S) scattering phase function [33, 41], the Fournier-Forand (F-F) phase function [42], the One-Term Henyey-Greenstein (OTHG) phase function [43], Haltrin's Two-Term Henyey-Greenstein (TTHG) phase function [44] and Petzold's average phase function [40, 45] in the range of 0.1 to 5°. The parameters of the approximately linear logarithmic scattering phase function (proposed in this paper), the S-S scattering phase function and the F-F scattering phase function are: the slope of the particle scale distribution $\xi = 3.4586$ and the refractive index of the main body of the particle $n = 1.16$. The root mean square error (RMSE) rate between the scattering phase functions $s_1(\theta)$ and $s_2(\theta)$ is defined as:

$$RMSE = \left\{ \frac{1}{K} \sum_{i=1}^{K} \left[ \frac{s_1(\theta) - s_2(\theta)}{0.5 \cdot [s_1(\theta) + s_2(\theta)]} \right]^2 \right\}^{0.5} \qquad (17)$$

where K is the number of samples of phase function measured by Petzold from 0.1° to 5°. The root mean square error rate between each scattering phase function and Petzold's average particle phase function is calculated in Table 1. The RMS of the S-S scattering phase function and the linear approximate logarithmic phase function of this paper are nearly equal, which is about 0.26. By comparison, the fitting accuracy of the proposed approximate linear logarithmic scattering phase function, S-S scattering phase function and F-F scattering phase function is much better than that of OTHG and TTHG scattering phase function in the range of 0.1° to 5°. The scattering phase function proposed in this paper can not only approximate the S-S scattering phase function well to achieve a good fitting effect but also has a simpler mathematical form than the S-S scattering phase function and F-F scattering phase function, which is more conducive to practical derivation and engineering applications.

**Table 1. The root mean square error of different phase functions**

| Angular ranges | phase functions | | | | |
|---|---|---|---|---|---|
| | Present | S-S | F-F | OTHG | TTHG |
| 0.1°-5° | 0.2612 | 0.2635 | 0.1241 | 1.3594 | 0.6383 |

**Funding**

The National Natural Science Foundation of China (NSFC) NO.61991454，CAS

Interdisciplinary Innovation Team Project Grant

**Disclosures**

The authors declare no conflicts of interest.